\documentclass[11pt]{article}
\usepackage{pstricks}
\usepackage{ctable}
\usepackage{multirow}
\usepackage{amsmath}

\makeatletter
\def\section{\@startsection{section}{1}{\z@}{1ex}{1ex}{\bf \large}}
\makeatother

\newfont{\bbbold}{msbm10}

\def\cN{{\cal N}}
\def\cO{{\cal O}}

\newfont{\goth}{eufm10 scaled \magstep1}

\def\gs{\mbox{\goth s}}

\def\gu{\mbox{\goth u}}

\def\a{\alpha}\def\adt{\dot \alpha}
\def\b{\beta}\def\bdt{\dot \beta}
\def\c{\gamma}\def\cdt{\dot\gamma}
\def\d{\delta}\def\ddt{\dot\delta}
\def\ve{\varepsilon}

\def\l{\lambda}

\def\th{\theta}

\let\la=\label

\def\nn{\nonumber}
\def\bd{\begin{document}}
\def\ed{\end{document}}
\def\be{\begin{equation}}
\def\ee{\end{equation}}
\def\ba{\begin{array}}
\def\ea{\end{array}}
\def\bea{\begin{eqnarray}}
\def\eea{\end{eqnarray}}
\def\ft#1#2{\tfrac{#1}{#2}}
\def\fft#1#2{\frac{#1}{#2}}
\def\sst#1{{\scriptscriptstyle #1}}
\def\oneone{\rlap 1\mkern4mu{\rm l}}

\newcommand{\eq}[1]{(\ref{#1})}
\newcommand{\w}[1]{\\[0.#1cm]}
\def\eqs#1#2{(\ref{#1}-\ref{#2})}
\def\det{{\rm det\,}}
\def\tr{{\rm tr}}
\def\del{\partial}
\newcommand{\hoch}[1]{$\, ^{#1}$}
\newcommand{\imperial}{\it\small Theoretical Physics Group, Imperial College London\\ Prince Consort Road, London SW7 2AZ, UK}
\newcommand{\kings}
{\it\small Department of Mathematics, King's College, University of London\\ Strand, London WC2R 2LS, UK}
\newcommand{\uu}
{\it\small Department of Theoretical Physics, Uppsala, Sweden}
\newcommand{\golm}
{\it\small AEI, Max Planck Institut f\"ur Gravitationsphysik\\ Am M\"{u}hlenberg 1, D-14476 Potsdam, Germany}
\makeatletter
\newcommand{\lapp}
{\it\small LAPTH, Universit{'e} de Savoie, CNRS, B.P. 110, F-74941 Annecy-le-Vieux Cedex France}
\newcommand{\dur}
{\it\small Institute for Particle Physics Phenomenology, Department of Mathematical Sciences and Department of Physics, Durham University, Durham DH1 3LE UK}
\@addtoreset{equation}{section} \makeatother

\newcommand{\auth}{\large J.M. Drummond\hoch{1}, P.J. Heslop\hoch{2} and P.S.\ Howe\hoch{3}}
\begin{document}

\renewcommand{\thefootnote}{\fnsymbol{footnote}}

\null
\begin{flushright}
%{\small AEI-2009-087}\\
{\small KCL-MTH-10-07}\\
{\small IPPP/10/71}\\
{\small DCTP/10/142}\\
{\small LAPP-035/10}
\vskip 1.5 cm
\end{flushright}

\begin{center}
{\Large{\bf  A note on $N=8$ counterterms}}
\vspace{.75cm}

\auth

\vspace{.5cm}

\begin{itemize}
\item [$^1$]\lapp\item[$^2$] \dur \item [$^3$] \kings
\end{itemize}
\vspace{1cm}

%%%%%%%%%%%%%%%%%%%%%%%%%
{\bf Abstract}
%%%%%%%%%%%%%%%%%%%%%%%%%%
\end{center}
\vskip .5cm
The most difficult counterterms to construct in any supersymmetric theory are those that cannot be written as full superspace integrals of gauge-invariant integrands. In $D=4$ maximal supergravity it has been known for some time that there are just three of these at the linearised level. In this article we discuss these counterterms again from the point of view of representations of the superconformal group. In particular, we show that the only independent invariants constructed from shortened superconformal multiplets in $D=4$ are BPS. 

\vspace{1cm}

\hrule

{\sl email: drummond@lapp.in2p3.fr; paul.heslop@durham.ac.uk; paul.howe@kcl.ac.uk}

\null

\renewcommand{\thefootnote}{\arabic{footnote}}
\setcounter{footnote}{0}

\pagebreak
%\tableofcontents
\setcounter{page}{1}

%%%%%%%%%%%%%%%%%%%%%%%%%%%%%%%%%%%%%%%%%%%%%%%%%%%%%%%%%%%%%%%%%%%%%%%%%%%%%%%%%%%%%%
%{\bf Introduction.}
\section*{Introduction}
%%%%%%%%%%%%%%%%%%%%%%%%%%%%%%%%%%%%%%%%%%%%%%%%%%%%%%%%%%%%%%%%%%%%%
Higher-order invariants in supersymmetric theories are important as possible field-theoretic counterterms and as higher-derivative terms in effective actions for strings or branes. These invariants fall into two classes, which we might call long and short, and which generalise D and F terms in $N=1, D=4$ supersymmetry. A D, or long, term, corresponds to an integral over the full superspace of some gauge-invariant superfield, while an F, or short, term is a chiral integral in $N=1$ or a suitable generalisation for $N>1$. There are many D terms but rather few F terms. In fact, in $D=4, N=8$ supergravity it was shown in~\cite{Drummond:2003ex} that there are only three of these in the linearised theory, each with four points ( $d^{2k}R^4$ integrals in spacetime with $k=0,2,3$.)

The existence of the three-loop $R^4$ counterterm was established (at
the linearised level) for $N=1,2$ supersymmetry in \cite{Deser:1977nt,Deser:1978br}, where they are
D-type invariants. In $N=8$ on the other hand, the simplest D-type counterterm 
does not occur until seven loops \cite{Howe:1980th}, and the $R^4$ invariant 
turns out to be a short F-term invariant. It was first constructed in~\cite{Kallosh:1980fi}, a manifestly covariant formulation making use of the notion of a superaction was given in~\cite{Howe:1981xy}, and it was shown in~\cite{Hartwell:1994rp} that this can be written very simply in harmonic superspace.  In~\cite{Drummond:2003ex} a study was made of all possible counterterms in harmonic superspaces and it was found that there are only two other F-terms, corresponding to counterterms at the five and six loop levels  ($d^4 R^4$ and $d^6 R^4$).  It is now known that $d=4, N=8$ supergravity is finite at three loops \cite{Bern:2007hh}, and more recently it has been shown that maximal supergravity is finite at four  loops in $D=5$ \cite{Bern:2009kd}, where the relevant invariant is $d^6 R^4$. Although the $D=4,N=8$ case has not been explicitly checked at five and six loop order, it would no longer be regarded as a surprise if it turned out to be finite here as well. 

In this article we revisit the $D=4, N=8$ F-term invariants, but from
a slightly different point of view to that adopted
in~\cite{Drummond:2003ex}. We shall still make use of the fact that
the linearised field strength superfield is superconformal, but we
avoid the use of harmonic superspace. Instead we approach the problem
by directly determining which superconformal representations possess
suitable singlet top components. Such top components automatically
give rise to supersymmetric integrals in spacetime. In particular,
this enables us to confirm very easily that there are no 
F-term invariants other than the three we have referred to above. More
generally, we show that there are no independent superactions with
measures that are not Lorentz scalars for massless supersymmetric
theories in $D=4$.

Before going into the details, we would like to comment on two
features of our analysis: the use of on-shell supersymmetry and the
linearisation of supergravity. No off-shell formulations are known for
either maximal super Yang-Mills (SYM) or  maximal supergravity. In
order to construct counterterms or higher-order terms in effective
actions it is therefore necessary to start with the on-shell
supersymmetry transformations of the original,  lowest-order  Lagrangian. The
addition of any on-shell deformation to the original action will
then be invariant up to terms proportional to the original equations
of motion. This procedure is perfectly satisfactory since such terms
can then be compensated by amending the supersymmetry
transformations. This will in turn induce higher-order terms in
the action, and the iteration of this procedure gives rise to a
perturbative method for handling the modified supersymmetry
transformations. An example of this is provided by the $F^4$ invariant
in $D=10$ SYM. In the abelian case this gives rise to an $F^6$
contribution at the next order and eventually to the full Born-Infeld series of terms. In the non-abelian case there is a similar single-trace $F^4$ deformation that gives rise to many more terms than just $F^6$ at the next order \cite{Koerber:2002zb}, a result that has recently been confirmed using only supersymmetry \cite{Howe:2010nu}. Of course, one would also want to know that the modified supersymmetry transformations do indeed satisfy the supersymmetry algebra. A convenient way of doing this, particularly relevant in the quantum-mechanical context, is to introduce ghosts and make use of the full BRST/BV formalism, see, for example, \cite{Dixon:1990jv,Howe:1990pz,Drummond:2003ex,Baulieu:2006gx}. It is possible there might be an obstruction to the extension of a deformation to all orders, which would indicate a supersymmetry anomaly of an unusual type, although this would be unexpected from a string theory perspective.

In the case of maximal supergravity we are also forced to deal with
linearised superfields when looking at possible F-terms. This is
legitimate for making comparisons with graviton scattering amplitudes,
but one would also eventually like to understand what the full
non-linear expressions are that correspond to the linearised ones. In
supergravity, therefore, we have to cope not only with the
non-linearities induced by the on-shell nature of the supersymmetry
transformations, but also the non-linearities of the full classical
theory. It is not easy to see how to generalise arbitrary F-terms to
the full theory because the linearised superspace measures do not have
obvious non-linear counterparts. Indeed, it could be the case that
some, or even all, of these invariants do not admit non-linear
extensions, and this might be the reason for the unexpected finiteness
results. It has been shown that $E_7$ symmetry can be maintained
in perturbation theory \cite{Bossard:2010dq}, at the cost of manifest Lorentz invariance, and this would be a further constraint that
would need to be satisfied. In a recent paper a string theory based argument has been given which shows that the full $R^4$ invariant is not $E_7$ invariant \cite{Elvang:2010kc}.\footnote{$E_7$ symmetry has also been invoked as a constraint in the context of light-cone superfields 
\cite{Kallosh:2009db,Kallosh:2010mk}}

In this note we shall not discuss these important issues
further; instead we restrict our 
discussion to the linearised level and use on-shell
supersymmetry. 
In this way we can be certain that we have all the allowed
allowed invariants 
(i.e. we have not missed any) although there remains the possibility that some of them
will not extend to genuine 
invariants of the full non-linear theory.

%%%%%%%%%%%%%%%%%%%%%%%%%%%%%%%%%%%%%%%%%%%%%%%%%%%%%%%%%%%%%%%%%%
%{\bf Linearised $N=8$ supergravity.}
\section*{ Linearised $N=8$ supergravity.}
%%%%%%%%%%%%%%%%%%%%%%%%%%%%%%%%%%%%%%%%%%%%%%%%%%%%%%%%%%%%%%%%%%
The spectrum of supergravity consists of the graviton, 8 gravitinos, 28 vector fields, 56 spin one-half fermions and 70 scalars. The whole set of component field strengths can be assembled into an $N=8$ scalar superfield $W_{ijkl},\ i,j=1\dots 8,$ that transforms under the 70-dimensional representation of $SU(8)$. It is therefore totally antisymmetric and self-dual. It depends on $x^a, a=0,1,2,3$ and 8 two-component fermionic coordinates and their complex conjugates $(\th^{\a i},\bar\th^{\adt}_i)$. $W_{ijkl}$ lives in flat $N=8$ superspace which is equipped with the supersymmetric invariant derivatives $(\del_a,D_{\a i},\bar D_{\adt}^i)$ where

\be
[D_{\a i},\bar D_{\bdt}^j]=i \d_i^j \del_{\a\bdt}
\la{2.1}
\ee

is the only non-trivial graded commutator, and where we have replaced the vector index on the spacetime derivative  by a pair of spinor indices. The superfield $W_{ijkl}$ is constrained to satisfy

\bea
D_{\a i} W_{jklm}&=&D_{\a [i} W_{jklm]}\nn\w1
\bar D_{\adt}^i W_{jklm}&=&-\frac{4}{5}\d^i_{[j} \bar D_{\adt}^n W_{klm]n}\nn\w1
\bar W^{ijkl}&=&\frac{1}{4!} \ve^{ijklmnpq} W_{mnpq}\ .
\la{2.2}
\eea

The third of these is the $SU(8)$ self-duality condition; it implies that the
first two are equivalent under complex conjugation. The differential constraints may be
interpreted as stating that the supersymmetry variation of the scalars
gives 56 spin one-half fields, i.e. the physical fields at this
level. One can easily check that the remaining independent components
of $W_{ijkl}$ are the field strengths of the fields listed above, for
example, the graviton field strength is the (linearised) Weyl
tensor, $C_{\a\b\c\d}\sim D_{\a i}\ldots D_{\d l}\bar W^{ijkl}$. All of the component fields obey the free-field equations of
motion, which is necessary in order for the number of bosonic and
fermionic degrees of freedom to match. 

A third important feature of our analysis is the use
of superconformal representation theory. At first
sight this might seem strange since maximal supergravity is certainly not
superconformal and we are interested in objects invariant under
supersymmetry, not superconformal symmetry, so we briefly explain
why we can do this.

Firstly, the component-field equations derived from (\ref{2.2}) are all conformal, since all of the fields
are massless (and free) while  $W_{ijkl}$ itself
transforms as a primary superfield  under the superconformal group
$SU(2,2|8)$.

Secondly, as in $N=1$ supersymmetry, any invariant, whether of D- or F- term
type, comes from taking the top component of a supermultiplet and
integrating it over space-time. 
Now in the $N=8$  case any such supermultiplet will be constructed from products of
$W$s and (super)derivatives acting on 
$W$s and hence transforms in a well-defined manner under the superconformal
group (which can be easily determined from the group's action on
$W_{ijkl}$). Any such supermultiplet can be decomposed into irreducible supermultiplets that transform under primary or descendant representations of the superconformal group.
Since a descendant supermultiplet will have the same top component as the primary from which it is descended (or is a spacetime derivative in which case it can be ignored in an integral), it follows that we need only consider top components that are contained in primary representations.

Therefore,  in order to classify the possible integral
invariants, we only need to classify all of the possible primaries
that can be constructed from $W_{ijkl}$ and determine which ones can contain suitable top components. 
The short invariants will come from short primary multiplets, or atypical superconformal representations, while the long invariants, D terms, correspond to typical representations which are also singlets under the Lorentz group and $SU(8)$, i.e. unconstrained scalar superfields.
Reducing the problem of finding invariants to that of finding suitable
superconformal representations is  extremely useful since there is much known about the
classification of superconformal representations which we can
straightforwardly exploit. We review this now.

%%%%%%%%%%%%%%%%%%%%%%%%%%%%%%%%%%%%%%%%%%%%%%%%%%%%%%%%%%%%%%%%%%%%%
%{\bf Superconformal representations.}
\section*{ Superconformal representations.}
%%%%%%%%%%%%%%%%%%%%%%%%%%%%%%%%%%%%%%%%%%%%%%%%%%%%%%%%%%%%%%

The generators of the superconformal algebra of $SU(2,2|N)$ are $(L,R, K,M,N\,|\,Q,S)$, corresponding to dilations, $U(1)$ R-symmetry, conformal boosts, Lorentz transformations, $SU(N)$, supersymmetry and special supersymmetry respectively. Representations of $N$-extended superconformal symmetry in $D=4$ are specified by $N+3$ quantum numbers $(L,R,J_1,J_2,a_1,\ldots
a_{N-1})$, where $L$ is the dilation weight, $R$ is the R-charge (for
these we
use the same labels for both the charge and the generator hopefully
without confusion),  $J_1$ $J_2$ are the two spin quantum numbers, and the $a_i$s are the Dynkin labels of an irreducible internal $SU(N)$ representation \cite{screp}. The unitary representations have to satisfy certain unitarity bounds which can be one of three types:

 \be
 \ba{rrclrcl}
{\rm Series\ A}:& L&\geq&2+2J_2-R+{2m \over N}, &L&\geq& 2 +
2J_1+R+2m_1-{2m\over N}\\
&&&&&&\nn\\
{\rm Series\ B}:& L&=&-R +{2m \over N}, &L &\geq&1+m_1 +J_1,
\qquad J_2=0\\
&&&&&&\nn\\
{\rm or}:& L&=&R +2m_1 -{2m \over N}, &L &\geq&1+m_1 +J_2,
\qquad J_1=0\\
&&&&&&\nn\\
{\rm Series\ C}:& L&=&m_1, &R&=&{2m \over N}-m_1, \qquad J_1=J_2=0
\ea\label{abc} 
\ee

Here $m=\sum ka_k$ is the total number of boxes in the Young tableau
corresponding to the $SU(N)$ representation $(a_1,\ldots a_{N-1})$, and
$m_1=\sum a_k$ is the number of boxes in the first row. Representations in series B and C are always short, while there are some series A representations that are short, although they are not BPS. 

These representations are all of highest weight type, and all the states in a given irreducible module are determined from the highest weight state by operating on it with the lowering operators. The highest weight, $\cO$ say, is annihilated by all $K$ and $S$ operators, and we can ignore these from now on. In addition, the dilation operator just gives the dimension of each state which we need not bother with since we know this from the dimension of the highest weight. The momentum operator $P_a$ essentially generates the spacetime-dependence of the module; we can instead replace the highest weight state by a highest weight field depending on $x$, which we shall also denote by $\cO$. The supersymmetry operators, $Q,\bar Q$ generate all  of the components of the superfield associated with the module, while the internal symmetry operators generate the internal symmetry modules at each level in the supersymmetry generators. A top component of a multiplet is one that is  annihilated by all $Q$s and $\bar Q$s up to a total derivative. Since we are interested in integrating top components over spacetime, we can, when operating on $\cO$ with the supersymmetry operators, assume that they all anticommute, since any spacetime derivative terms will integrate to zero.

In order to understand the irreducible representations more easily we will
construct any representation $\cO$ out of three building-block
representations  (see~\cite{Ferrara:2000eb,Heslop:2000mr} for a similar approach)

\begin{equation}
  \cO = \cO^{(1)} \, \cO^{(C)} \, \cO^{(2)} \ .
\end{equation}
$\cO^{(1)}$ will provide the left-handed spin quantum number $J_1$,
$\cO^{(2)}$ the right-handed spin quantum number $J_2$ and
$\cO^{(C)}$ the $SU(N)$ quantum numbers. Any shortening conditions on
$\cO$ will be a consequence of the shortening conditions of the
building block representations which we will give shortly. For series
B representations one of 
either $\cO^{(1)}$ or $\cO^{(2)}$ will be trivial, i.e. 1, and for series C
representations both $\cO^{(1)}$ and $\cO^{(2)}$ will be trivial
leaving $\cO=\cO^{(C)}$. The building
blocks for any given irreducible representation are unique and all irreducible representations can be so constructed.

The building block $\cO^{(C)}$, if non-trivial, has non-zero quantum numbers, $a_1 \dots
a_{N-1}$ together with $L=m_1$ and $R=2m/N-m_1$. It corresponds to a series C representation. The
highest weight state of such  a representation is
annihilated by the $\gs\gu(N)$ raising operators $N_i{}^j,\,j>i$. It
is also annihilated by (both components of) $Q_r, r=1 \ldots p$ and
$\bar Q^{r'},r'=(N-q)\ldots N$, where $a_p$ is the left-most non-zero
$\gs\gu(N)$ label and $a_{N-q}$ the right-most non-zero one.
It is not
difficult to see that these constraints are consistent with this
state's being annihilated by the $\gs\gu(N)$ raising
operators. Moreover, this set of $Q$s and $\bar Q$s is anticommutative
so that no spacetime constraint is generated.

 The building block $\cO^{(1)}$ has non-zero quantum numbers $J_1$ and
 $L=-R\geq 1+J_1$. It is chiral, i.e. annihilated by all of the $\bar
 Q$s, and if the bound on the dilation weight $L$ is saturated it obeys a further $Q$
 constraint (a divergence constraint if $J_1>0$ or a second-order
 constraint if $J_1=0$). In this case $\cO^{(1)}$
 corresponds to an on-shell chiral massless multiplet. 
The building block $\cO^{(2)}$ is the conjugate of an $\cO^{(1)}$
representation for 
which the $Q$ and $\bar Q$ constraints are interchanged. All of this
is summarised in the table.

\vskip .5cm

 \begin{table}[h!]
   \centering
   \begin{tabular}{lcc}\toprule
&non-vanishing quantum numbers& $Q,\bar Q$ constraints\\\toprule
$\cO^{(C)}$&$
\begin{array}{c}
  L=m_1,\quad R=2m/N-m_1\\
a_p,\ldots, a_{N-q}
\end{array}
$&
$\begin{array}{rcl}
Q_r \cO^{(C)} &=&0, \qquad r=1\ldots p\\
\bar Q^{r'} \cO^{(C)} &=&0, \qquad r'=N-q \ldots N-1     
\end{array}$\\\midrule[0.08em]
$\cO^{(1)}_{\rm{long}}$&
$\begin{array}{c}
  L=-R>1+J_1,\quad J_1
\end{array}$
&
$\begin{array}{rcl}
\bar Q^i \cO^{(1)} &=&0
\end{array}$\\\midrule
$\cO^{(1)}_{\rm{short}}$&
$\begin{array}{c}
  L=-R=1+J_1,\quad J_1\neq 0
\end{array}$
&
$\begin{array}{rcl}
\bar Q^i \cO^{(1)} &=&0 \qquad
Q^\a_i \cO^{(1)}_{\a\ldots}=0
\end{array}$\\\midrule
$\cO^{(1)}_{\rm{short}}$&
$\begin{array}{c}
  L=-R=1,
\end{array}$
&
$\begin{array}{rcl}
\bar Q^i \cO^{(1)} &=&0\qquad 
Q^2_{ij} \cO^{(1)}=0
\end{array}$\\\midrule[0.08em]
$\cO^{(2)}_{\rm{long}}$&
$\begin{array}{c}
  L=R>1+J_2,\quad J_2
\end{array}$
&
$\begin{array}{rcl}
Q_i \cO^{(2)} &=&0
\end{array}$\\\midrule
$\cO^{(2)}_{\rm{short}}$&
$\begin{array}{c}
  L=R=1+J_2,\quad J_2\neq 0
\end{array}$
&
$\begin{array}{rcl}
Q_i \cO^{(2)} &=&0 \qquad
\bar Q ^{\adt i} \cO^{(2)}_{\adt\ldots}=0
\end{array}$\\\midrule
$\cO^{(2)}_{\rm{short}}$&
$\begin{array}{c}
  L=R=1,
\end{array}$
&
$\begin{array}{rcl}
 Q_i \cO^{(2)} &=&0\qquad 
(\bar Q^2)^{ij} \cO^{(2)}=0
\end{array}$\\\bottomrule
   \end{tabular}
   \caption{Summary of the quantum numbers (only those which may be
     non-zero) together with the shortening conditions of the building
     block representations, $\cO^{(1)}, \cO^{(C)}$ and
     $\cO^{(2)}$. One can easily see from these quantum numbers that the quantum number of the
     representation $\cO = \cO^{(1)} \, \cO^{(C)} \, \cO^{(2)}$ will
     satisfy both series A bounds~\eq{abc}. Similarly  $\cO =
     \cO^{(C)} \, \cO^{(2)}$ and  $\cO = \cO^{(1)} \, \cO^{(C)}$  are
     series B representations and  $\cO = \cO^{(C)}$ is a series C
     representation. $Q^2_{ij}:= Q^\a_i Q_{\a j}=Q^2_{ji}$}  
   \label{tab}
 \end{table}

%%%%%%%%%%%%%%%%%%%%%%%%%%%%%%%%%%%%%%%%%

\section*{Invariants from irreducible representations}

%%%%%%%%%%%%%%%%%%%%%%%%%%%%%%%%%%%%%%%%%

In order to find invariants we need to find
multiplets whose top components are both Lorentz scalars and $SU(N)$
singlets. In this section
we will rule out certain classes of representations on the grounds that
they  cannot have scalar top components.
Specifically, any representation which contains either a chiral or
anti-chiral massless multiplet,  $\cO^{(1)}_{\rm
  short}$ or $\cO^{(2)}_{\rm
  short}$, as one of its building blocks, can never have a scalar top component.

To prove this, consider first a representation 
constructed as $\cO=\cO^{(1)}_{\rm short}\cO^{(C)} \cO^{(2)}$ where
$\cO^{(2)}$ can be short or long (or indeed trivial). The $Q$-constraints on this
operator can be read off from the constraints on the building block
operators from the table.
If the
representation is a scalar, so that $J_1=0$, the highest weight
will satisfy a second-order constraint with respect to the subset of
the $Q$s that annihilate the series C representation, $Q^2_{r s}
\cO=0,\quad r,s=1\ldots p$. If the representation has non-zero spin
on the other hand, the highest weight state will satisfy a
divergence constraint with respect to the subset of $Q$s that are
annihilated by the series C centre, $Q^\alpha_{r}
\cO_{\alpha\b\ldots}=0, \quad r,s=1\ldots p$.  

We can now prove very easily that there are no invariants that can be
constructed from operators which have $\cO^{(1)}_{\rm short}$ as a
building block (and by conjugation therefore the same is true for  operators which have $\cO^{(2)}_{\rm short}$ as a
building block).
These correspond to series A and B representations which 
saturate one or both of the unitarity bounds.

This assertion rests on the
simple fact that the top components of such multiplets cannot be Lorentz
scalars. 
To see this it is enough to  consider the top $Q$-component (i.e. the component
obtained by applying as
many $Q$s as possible to the highest weight state) of
$\cO^{(1)}_{\rm{short}} \cO^{(C)} \cO^{(2)}$ where $\cO^{(2)}$ is kept
general (although it is anti-chiral, i.e. annihilated by all the $Q$s).
On considering the various constraints, it is clear that the top $Q$-component
has left spin $J_1+p/2$ and is explicitly given by

\begin{equation}
 Q^{(\alpha_1}_1\dots Q^{\alpha_p}_p [Q_{p+1}\dots Q_N]^2\ 
 \left(\cO^{(1)\,\alpha_{p+1} \dots \alpha_{p+2J_1)}} \  \cO^{(C)} \cO^{(2)}\right) \ .
\end{equation}

Since $p\geq 1$ it immediately follows that the
top component of any multiplet involving $\cO^{(1)}_{\rm{short}}$ can
never be a Lorentz scalar.\footnote{Note that to keep things simple we have
kept $\cO^{(2)}$ general and 
have ignored the action of $\bar Q$. Since dotted indices can not be
contracted with 
undotted indices the top component has the same left spin as the top
$Q$-component. } 
Similar conclusions regarding the top components of series
A and B representations in the case of $\cN$=4 SYM can
be read off from the results of ~\cite{Dolan:2002zh}.

We conclude that there can be no scalar top components of series A or
B representations, saturating a unitary bound. The only remaining
possibilities are therefore $\cO=\cO^{(1)}_{\rm
  long}\cO^{(C)}\cO^{(2)}_{\rm long}$ (i.e. long series A
representations) $\cO=\cO^{(C)}\cO^{(2)}_{\rm long}$ or $\cO=\cO^{(1)}_{\rm
  long}\cO^{(C)}$ ``long'' series B representations,   or
$\cO=\cO^{(C)}$ series C representations.

The  series C
representations and long series A representations can give rise to
invariants; the series C case was
analysed in a number of different theories including $N=8$
in~\cite{Drummond:2003ex}.

The simplest representations to consider from the point of view of
integral invariants are the long multiplets,  corresponding to series A
reps with  the unitary bound unsaturated.  The highest weight field
$\cO$ must be a Lorentz and $SU(N)$ scalar, satisfying no supersymmetry constraints, and
then the top component is given by $[Q_1\ldots Q_N\,\bar Q^1\ldots
\bar Q^N]^2\cO$ (each $Q$ and $\bar Q$ is a two-component
spinor). This is also a Lorentz and $SU(N)$ scalar, and hence defines
a supersymmetric invariant integral. It may carry dilation and R
weights, but this does not spoil things. There will  be an infinite
number of such long invariants and in the $N=8$ context they first
appear at seven loops \cite{Howe:1980th}.

 As a simple example of series C consider the multiplet in $N=2$ with non-zero quantum numbers $L=4, a_1=4$, i.e. the only non-zero super-Dynkin label is $n_3=4$. This corresponds to a one-half BPS scalar superfield $\cO_{ijkl}$ in the 5 of $SU(2)$. The highest weight field is $\cO_{1111}(x)$; it is annihilated by $Q_1$ and $\bar Q^2$. The top component of this multiplet  is therefore $[Q_2\bar Q^1]^2 \cO_{1111}$. This clearly determines a supersymmetric invariant since operating on this with either $Q_1$ or $\bar Q^2$ is zero up to a spacetime derivative. Notice that this expression is invariant under $\gs\gu(2)$, even though each factor has a particular numerical index. The raising operator $N_1{}^2$ annihilates $\cO_{1111}$,  and commuted with $Q_2$ or $\bar Q^1$ gives either $Q_1$ or $\bar Q^2$ both of which also annihilate $\cO_{1111}$. Moreover, it has zero charge under the $\gu(1)$ sub-algebra of $\gs\gu(2)$. Since it is a highest weight state under $\gs\gu(2)$ and has charge zero, it must be a singlet and therefore annihilated by $N_2{}^1$. The fact that the top component is an internal singlet is important. There is an infinite number of one-half BPS supermultiplets, but only the one with $a_1=4$ gives rise to an $SU(2)$-invariant supersymmetric integral.

%%%%%%%%%%%%%%%%%%%%%%%%%%%%%%%%%%%%%%%%%%%%%%%%%%%%%%%%%%%%%%%%%%%%%
%{\bf $N=8$ invariants.}
\section*{ $N=8$ invariants.}
%%%%%%%%%%%%%%%%%%%%%%%%%%%%%%%%%%%%%%%%%%%%%%%%%%%%%%%%%%%%%%%%

Now let us focus on the theory of interest, $N=8$ supergravity. First we consider the series C multiplets, for which the highest weight state is annihilated by some consistent subset of the supersymmetry generators. The field strength itself is one of these, with $L=1$ and $a_4=1$ being the only non-zero quantum numbers. The highest weight is $\cO_{1234}=W_{1234}(\th=0)$. The shortest series C representations are one-half BPS for which the highest weight state is annihilated by half of the supersymmetry generators. One can see that this set has to consist of four $Q$s and four $\bar Q$s owing to the fact that the candidate BPS multiplets must be products of $W$s. Multiplets of this sort therefore have highest weight states of the form $\cO= (W_{1234})^p$ for some $p$, and the top component of such a multiplet is $[Q_5\ldots Q_8\bar Q^1\ldots\bar Q^4]^2 \cO$, for $p\geq 4$. There is clearly only one choice of $p$, $p=4$, for which this top component is a singlet. This is the three-loop $R^4$ invariant. Our construction is rather similar to the original one \cite{Kallosh:1980fi}, but makes it clearer that the integral is $SU(8)$ invariant. One can investigate systematically all the other possibilities. This was done in~\cite{Drummond:2003ex}  and we do not repeat it here. There are just two, $d^4 R^4$ and $d^6 R^4$, with highest weight states that are annihilated by two $Q$s and $\bar Q$s and one $Q$ and one $\bar Q$ respectively.

As an example of a vanishing theorem, we prove that there are no four loop invariants. The only possibility is a highest weight state that is annihilated by $Q_r, r=1,2,3$ and $\bar Q^{s'},s'=6,7,8$. The top component will therefore be $[Q_4\ldots Q_8\bar Q^1\ldots \bar Q^5]^2 \cO$. In order for this to be a singlet $\cO$ must be of the form $\cO_{1111222233334455}$ (corresponding to the $\gs\gu(8)$ Dynkin labels $(0,0,2,0,2,0,0)$). This would have to be $(W_{1234})^2 (W_{1235})^2$, but it is easy to see that this is not a highest weight state because it is not annihilated by $N_4{}^5$. This explicit example also makes it clear why the highest weight state for any putative BPS invariant can only have four fields. The state must be annihilated by at least one $Q$, say $Q_1$, and at least one $\bar Q$, $\bar Q^8$ say, and this means that $\cO$ must have exactly four 1s, together with various other numerical indices.\footnote{Up to 7; the index 8 can be eliminated as $W$ transforms under an $SU(8)$ representation.} The only way of achieving this is to have four $W$s each having one 1 index. (A factor of $W$ without a 1 index is of course not annihilated by $Q_1$.)

We now turn to series A and B. As we have seen above, the only such
representations that can give rise to integral invariants are ``long''
series B. These are partially chiral. A study of the possible multiplets of this type for $N=8$ supergravity was made in \cite{Drummond:2003ex}. It was found that there is only one possibility, at three points, that has the right quantum numbers to be a candidate partially chiral integrand. However, it turns out that this multiplet satisfies the series B unitarity bound and so satisfies a second-order constraint as well as being partially chiral. It therefore integrates to zero. There are examples of series B multiplets that do not satisfy a second-order constraint, for example, one can simply take the square of this three-point candidate, but none of these have the right quantum numbers to be integrands. 

%%%%%%%%%%%%%%%%%%%%%%%%%%%%%%%%%%%%%%%%%%%%%%%%%%%%%%%%%%%%%%
\section*{The helicity structure of counterterms.}
%{\bf Self-dual amplitudes?}
%%%%%%%%%%%%%%%%%%%%%%%%%%%%%%%%%%%%%%%%%%%%%%%%%%%%%

A different proposal for classifying counterterms to that
initiated in~\cite{Drummond:2003ex} and developed further here was put forward in~\cite{Elvang:2010jv}.
It is  based on a study of the corresponding amplitudes. Here, we explicitly connect the two
approaches,  show how the helicity structure of
amplitudes is related to the particular superconformal operators we
consider, and give a very simple explanation for a bound on
the helicity structure of counterterms conjectured in reference~\cite{Elvang:2010jv} .

Let us now consider the helicity structure of the amplitudes the possible counterterms can contribute to. Let us recall that the on-shell $N=8$ supergravity multiplet is CPT self-conjugate and contains all particles from the negative helicity graviton (with helicity $-2$) to the positive helicity graviton (with helicity $2$). Writing it as an on-shell superfield  in light-cone momentum superspace  we have\footnote{ There is a simple relation between the light-cone chiral superfield and $W_{ijkl}$, see \cite{Bossard:2009sy}.}

\be
\Phi(\eta) = g^{++} + \eta^i \Gamma_i + \ldots + \eta^{i_1}\ldots \eta^{i_7} \epsilon_{i_1 \ldots i_8} \overline{\Gamma}^{i_8} + (\eta)^8 g^{--}.
\ee

Here $\eta^i$ is a Grassmann variable transforming in the fundamental representation of the $SU(8)$ R-symmetry group. We have included only the first two and last two terms in the superfield expansion for brevity. The expansion begins with the positive helicity graviton and gravitino and ends with the corresponding negative helicity states. The dots stand for all other on-shell states that appear between.

On-shell N=8 supersymmetry dictates that the sum of the helicities of the particles in a given amplitude must lie between $8-2n$ and $2n-8$. An amplitude with total helicity $8-2n+4k$ is called an N${}^k$MHV amplitude and when $k=n-4$ (its maximum value) the amplitude is often referred to as an $\overline{\rm MHV}$ amplitude.

The three F terms which arise as possible counterterms are all four-point invariants. As such the total helicity must be zero (and the amplitude is both MHV and $\overline{\rm MHV}$). The pure-graviton amplitudes which these counterterms contribute to therefore involve two negative and two positive helicity gravitons. Let us split the on-shell curvature tensor into its chiral and anti-chiral Weyl curvatures, $R=(C,\bar{C})$, containing the negative helicity graviton and positive helicity graviton respectively,

\be
C_{\a \b \c \d} = \l_\a \l_\b \l_\c \l_\d g^{++}, \qquad \bar{C}_{\adt \bdt \cdt \ddt} = \tilde{\l}_{\adt} \tilde{\l}_{\bdt} \tilde{\l}_{\cdt} \tilde{\l}_{\ddt} g^{--}.
\ee

Here we have introduced the spinor helicity variables describing the on-shell momentum of the particle $p^{\a \adt} = \l^\a \tilde{\l}^{\adt}$.
The three F terms thus have the pure-gravity structure, $C^2 \bar{C}^2$, $d^4 C^2 \bar{C}^2$ and $d^6 C^2 \bar{C}^2$, corresponding to the three-loop, five-loop and six-loop counterterms respectively.

When we consider long multiplets, many more counterterms can be constructed. The simplest suitable long multiplet is the one whose highest weight state is a product of four scalars. This four-point counterterm occurs at seven loops and has the pure gravity structure $d^8 C^2 \bar{C}^2$. As a full superspace integral it can be written

\be
\int d^{32} \theta W^4 = D^{16} \bar{D}^{16} W^4 =  d^8 C^2 \bar{C}^2 + \ldots
\ee

where the dots refer to terms involving fields other than the curvature.
As it is a four-point counterterm it is again of MHV type (and $\overline{\rm MHV}$ type). 

Let us generalise this example to other long multiplets. We will consider multiplets whose top components can be written
\be
D^p \bar{D}^q W^n = d^w C^u \bar{C}^v + \ldots
\ee
Here $n$ is the number of fields in the linearised counterterm. There must be at least 16 superderivatives of either chirality, $p\geq 16$, $q \geq 16$ as this is the top component of a long multiplet. Since the counterterm contains $v$ anti-chiral Weyl tensors $\bar{C}$ it contributes to N${}^{k}$MHV amplitudes for $k=v-2$. From counting the dimensions we see that it arises at $l=\frac{w}{2} + u+v-1$ loops.
Since $d=D\bar{D}$, $C = D^4 W$ and $\bar{C}=\bar{D}^4 W$ we can see that there is a simple relation between $(p,q,n)$ and $(w,u,v)$. Indeed we have

\begin{align}
u &= \tfrac{1}{8}(p-q+4n),\\
v &= \tfrac{1}{8}(q-p+4n),\\
w &= \tfrac{1}{2}(p+q-4n).
\end{align}
Thus we find the loop order $l=\frac{1}{4}(p+q) -1$ and the MHV degree $k=v-2=\frac{1}{8}(q-p+4n)-2$. Rearranging we find that
\be
k = \tfrac{1}{2}(q/2  +n -5 -l)
\ee
and since $q\geq16$ we find a bound on the chirality of a given counterterm,
\be
k \geq \frac{3+n-l}{2}.
\label{kbound}
\ee

By parity there is an equivalent bound coming from $p \geq 16$.  The bound \eq{kbound} agrees with that conjectured in \cite{Elvang:2010jv}.
Thus at a given number of points and a given loop order a counterterm can only violate helicity by a certain amount as dictated by the above bound. Note that this analysis straightforwardly rules out the seven loop counterterms of the type $d^6R^5$ and $d^2R^7$ since for any helicity configuration they would require fewer than 16 superderivatives of one or other type. The absence of these counterterms was shown in \cite{Elvang:2010jv} where explicit examples of non-vanishing MHV and NMHV matrix elements were also constructed corresponding to the long multiplets considered here. It is also simple to see that there will be a non-vanishing N${}^2$MHV counterterm of the type $R^8$. In general, beyond seven loops, one would expect that counterterms of any pure gravity type can exist as long as the helicity structure respects the bound (\ref{kbound}) and the corresponding parity conjugate bound.

We have not by any means given an exhaustive list of all possible long multiplets which give counterterms of a given MHV type at a given loop order. One could go on to count all possible ways such counterterms can be constructed, which amounts to counting all possible long multiplets with a given dimension, given total number of fields and a given chirality.  Counting  operators in superconformal theories can be done
using partition functions and supercharacters. This was carried out for $\cN$=4 super Yang-Mills in~\cite{Bianchi:2006ti},  and would be fairly  straightforward to generalise to $\cN$=8.

We emphasise that the above analysis holds for the linearised theory. In the full theory, full superspace integrals of terms involving arbitrary functions of the scalar fields are unlikely to be $E_7$ invariant.

%%%%%%%%%%%%%%%%%%%%%%%%%%%%%%%%%%%%%%%%%%%%%%%%%%%%%%%%%%%%%%
\section*{ Concluding remarks}
%{\bf  Concluding remarks}
%%%%%%%%%%%%%%%%%%%%%%%%%%%%%%%%%%%%%%%%%%%%%%%%%%%%%%%%%%%

In this note we have given a very simple discussion of the F terms that are allowed in $D=4, N=8$ supergravity based on the observation that the linearised field strength superfield is a superconformal primary field. The fact that there are only three such invariants was derived in our earlier paper \cite{Drummond:2003ex} but we did not give the details of the impossibility of constructing invariants from non-BPS short primaries there. In \cite{Elvang:2010jv}, along with other results, the fact that there are only three counterterms below the full superspace threshold , i.e. at six loops or fewer, was derived by a completely different method making use of scattering amplitude techniques. The argument given here is more general, as regards the F term issue, in that it shows that there are no independent non-Lorentz invariant superactions for any supersymmetric theory in $D=4$ built from multiplets that are superconformal, a category that includes super Yang-Mills theories, linearised supergravity and massless Wess-Zumino and hypermultiplets. 

This result does not necessarily hold in other dimensions. For example, consider $(1,1)$ supersymmetry in $D=6$. There the left and right chiral spinors can be contracted so a series A type supermultiplet could in principle have a scalar top component. 

We conclude by remarking on the significance of these counterterms for the ultra-violet properties of $N=8$ supergravity. It has been known for some years that the theory is three-loop finite \cite{Bern:2007hh}, a result that is in line with expectations from considerations in field theory  \cite{Howe:2002ui,Bossard:2009sy} and string theory \cite{Green:2006yu}. More recently, it has been established that the theory is finite at four loops \cite{Bern:2009kd} in $D=5$ where the relevant invariant is $d^6 R^4$. As we showed in \cite{Drummond:2003ex} the $D=4$ theory is finite at four loops, owing to the absence of an invariant.\footnote{See \cite{Kallosh:2009jb} for an alternative discussion of this point.} However, given the $D=5$ result it would not be a surprise if the $D=4$ theory turned out to be finite at five and six loops as well, even though there are linearised counterterms. One possible explanation for this could be, as we mentioned earlier, that the linearised counterterms do not admit duality invariant extensions in the full theory.

\vskip .5cm

%%%%%%%%%%%%%%%%%%%%%%%%%%%%%%%%%%%%%%%%%%%%%%%%%%%%%%%
\section*{Acknowledgements}

We thank the authors of reference \cite{Elvang:2010jv} for e-mail correspondence and Sven Kerstan for helpful comments. PSH thanks G. Bossard, U. Lindstrom, K. Stelle and L. Wulff for stimulating discussions.


\begin{thebibliography}{99}


%\cite{Drummond:2003ex}
\bibitem{Drummond:2003ex}
  J.~M.~Drummond, P.~J.~Heslop, P.~S.~Howe and S.~F.~Kerstan,
  ``Integral invariants in N = 4 SYM and the effective action for  coincident
  D-branes,''
  JHEP {\bf 0308} (2003) 016
  [arXiv:hep-th/0305202].
  %%CITATION = JHEPA,0308,016;%%

%\cite{Deser:1977nt}
\bibitem{Deser:1977nt}
  S.~Deser, J.~H.~Kay and K.~S.~Stelle,
  ``Renormalizability Properties Of Supergravity,''
  Phys.\ Rev.\ Lett.\  {\bf 38} (1977) 527.
  %%CITATION = PRLTA,38,527;%%
  
%\cite{Deser:1978br}
\bibitem{Deser:1978br}
  S.~Deser and J.~H.~Kay,
  ``Three Loop Counterterms For Extended Supergravity,''
  Phys.\ Lett.\  B {\bf 76} (1978) 400.
  %%CITATION = PHLTA,B76,400;%% 
  
%\cite{Howe:1980th}
\bibitem{Howe:1980th}
  P.~S.~Howe and U.~Lindstrom,
  ``Higher Order Invariants In Extended Supergravity,''
  Nucl.\ Phys.\  B {\bf 181} (1981) 487.
  %%CITATION = NUPHA,B181,487;%%   



\bibitem{Kallosh:1980fi}
  R.~E.~Kallosh,
  ``Counterterms in extended supergravities,''
  Phys.\ Lett.\  B {\bf 99} (1981) 122.
  %%CITATION = PHLTA,B99,122;%%



\bibitem{Howe:1981xy}
  P.~S.~Howe, K.~S.~Stelle and P.~K.~Townsend,
  ``Superactions,''
  Nucl.\ Phys.\  B {\bf 191} (1981) 445.
  %%CITATION = NUPHA,B191,445;%%
  
%\cite{Hartwell:1994rp}
\bibitem{Hartwell:1994rp}
  G.~G.~Hartwell and P.~S.~Howe,
  ``(N, P, Q) Harmonic Superspace,''
  Int.\ J.\ Mod.\ Phys.\  A {\bf 10} (1995) 3901
  [arXiv:hep-th/9412147].
  %%CITATION = IMPAE,A10,3901;%%
  
%\cite{Bern:2007hh}
\bibitem{Bern:2007hh}
  Z.~Bern, J.~J.~Carrasco, L.~J.~Dixon, H.~Johansson, D.~A.~Kosower and R.~Roiban,
  ``Three-Loop Superfiniteness of N=8 Supergravity,''
  Phys.\ Rev.\ Lett.\  {\bf 98} (2007) 161303
  [arXiv:hep-th/0702112].
  %%CITATION = PRLTA,98,161303;%%  
  
%\cite{Bern:2009kd}
\bibitem{Bern:2009kd}
  Z.~Bern, J.~J.~Carrasco, L.~J.~Dixon, H.~Johansson and R.~Roiban,
  ``The Ultraviolet Behavior of N=8 Supergravity at Four Loops,''
  Phys.\ Rev.\ Lett.\  {\bf 103} (2009) 081301
  [arXiv:0905.2326 [hep-th]].
  %%CITATION = PRLTA,103,081301;%%
  

  
%\cite{Koerber:2002zb}
\bibitem{Koerber:2002zb}
  P.~Koerber and A.~Sevrin,
  ``The non-abelian D-brane effective action through order $\alpha'^4$,''
  JHEP {\bf 0210} (2002) 046
  [arXiv:hep-th/0208044].
  %%CITATION = JHEPA,0210,046;%%
  
%\cite{Howe:2010nu}
\bibitem{Howe:2010nu}
  P.~S.~Howe, U.~Lindstrom and L.~Wulff,
  ``D=10 supersymmetric Yang-Mills theory at $\alpha'^4$,''
  JHEP {\bf 1007} (2010) 028
  [arXiv:1004.3466 [hep-th]].
  %%CITATION = JHEPA,1007,028;%% 
  
%\cite{Dixon:1990jv}
\bibitem{Dixon:1990jv}
  J.~A.~Dixon,
  ``Supersymmetry Is Full Of Holes,''
  Class.\ Quant.\ Grav.\  {\bf 7} (1990) 1511.
  %%CITATION = CQGRD,7,1511;%%
  
%\cite{Howe:1990pz}
\bibitem{Howe:1990pz}
  P.~S.~Howe, U.~Lindstrom and P.~White,
  ``Anomalies and renormalisation in the BRST-BV framework,''
  Phys.\ Lett.\  B {\bf 246} (1990) 430.
  %%CITATION = PHLTA,B246,430;%%
  
%\cite{Baulieu:2006gx}
\bibitem{Baulieu:2006gx}
  L.~Baulieu, G.~Bossard and S.~P.~Sorella,
  ``Shadow fields and local supersymmetric gauges,''
  Nucl.\ Phys.\  B {\bf 753} (2006) 273
  [arXiv:hep-th/0603248].
  %%CITATION = NUPHA,B753,273;%%
  
%\cite{Bossard:2010dq}
\bibitem{Bossard:2010dq}
  G.~Bossard, C.~Hillmann and H.~Nicolai,
  ``Perturbative quantum E7(7) symmetry in N=8 supergravity,''
  arXiv:1007.5472 [hep-th].
  %%CITATION = ARXIV:1007.5472;%%  
  
\bibitem{Elvang:2010kc}
  H.~Elvang and M.~Kiermaier,
  ``Stringy KLT relations, global symmetries, and $E_7(7)$ violation,''
  arXiv:1007.4813 [hep-th].
  %%CITATION = ARXIV:1007.4813;%%
  
 %\cite{Kallosh:2009db}
\bibitem{Kallosh:2009db}
  R.~Kallosh,
  ``N=8 Supergravity on the Light Cone,''
  Phys.\ Rev.\  D {\bf 80} (2009) 105022
  [arXiv:0903.4630 [hep-th]].
  %%CITATION = PHRVA,D80,105022;%%
  
%\cite{Kallosh:2010mk}
\bibitem{Kallosh:2010mk}
  R.~Kallosh and P.~Ramond,
  ``Light-by-Light Scattering Effect in Light-Cone Supergraphs,''
  arXiv:1006.4684 [hep-th].
  %%CITATION = ARXIV:1006.4684;%%   

       

\bibitem{screp}
M.Flato and C. Fronsdal,  Lett. Math. Phys. {\bf 8} (1984) 159;
V.K. Dobrev and V.B. Petkova, Phys. Lett. {\bf B162} (1985) 127,
Fortschr. Phys. {\bf 35} (1987) 537; B. Binegar, Phys. Rev. {\bf
D34} (1986) 525; B. Morel, A. Sciarrino and P. Sorba, Phys. Lett
{\bf B166} (1986) 69, erratum {\bf B167} (1986) 486.

\bibitem{Ferrara:2000eb}
  S.~Ferrara and E.~Sokatchev,
  ``Superconformal interpretation of BPS states in AdS geometries,''
  Int.\ J.\ Theor.\ Phys.\  {\bf 40} (2001) 935
  [arXiv:hep-th/0005151].
  %%CITATION = IJTPB,40,935;%%
  
\bibitem{Heslop:2000mr}
  P.~Heslop and P.~S.~Howe,
  ``Harmonic superspaces and superconformal fields,''
  arXiv:hep-th/0009217.
  %%CITATION = HEP-TH/0009217;%%
  
  
  

\bibitem{Dolan:2002zh}
  F.~A.~Dolan and H.~Osborn,
  ``On short and semi-short representations for four dimensional superconformal
  symmetry,''
  Annals Phys.\  {\bf 307} (2003) 41
  [arXiv:hep-th/0209056].
  %%CITATION = APNYA,307,41;%%
  
  
  
 \bibitem{Elvang:2010jv}
  H.~Elvang, D.~Z.~Freedman and M.~Kiermaier,
  ``A simple approach to counterterms in N=8 supergravity,''
  arXiv:1003.5018 [hep-th].
  %%CITATION = ARXIV:1003.5018;%%
  
%\cite{Elvang:2010jv}
\bibitem{Bianchi:2006ti}
 M.~Bianchi, F.~A.~Dolan, P.~J.~Heslop and H.~Osborn,
 ``N = 4 superconformal characters and partition functions,''
 Nucl.\ Phys.\  B {\bf 767} (2007) 163
 [arXiv:hep-th/0609179].
 %%CITATION = NUPHA,B767,163;%%
  
%\cite{Howe:2002ui}
\bibitem{Howe:2002ui}
  P.~S.~Howe and K.~S.~Stelle,
  ``Supersymmetry counterterms revisited,''
  Phys.\ Lett.\  B {\bf 554} (2003) 190
  [arXiv:hep-th/0211279].
  %%CITATION = PHLTA,B554,190;%%    
  
  
%\cite{Bossard:2009sy}
\bibitem{Bossard:2009sy}
  G.~Bossard, P.~S.~Howe and K.~S.~Stelle,
  ``The ultra-violet question in maximally supersymmetric field theories,''
  Gen.\ Rel.\ Grav.\  {\bf 41} (2009) 919
  [arXiv:0901.4661 [hep-th]].
  %%CITATION = GRGVA,41,919;%%  
  

  
%\cite{Green:2006yu}
\bibitem{Green:2006yu}
  M.~B.~Green, J.~G.~Russo and P.~Vanhove,
  ``Ultraviolet properties of maximal supergravity,''
  Phys.\ Rev.\ Lett.\  {\bf 98} (2007) 131602
  [arXiv:hep-th/0611273].
  %%CITATION = PRLTA,98,131602;%%   
  
%\cite{Kallosh:2009jb}
\bibitem{Kallosh:2009jb}
  R.~Kallosh,
  ``On UV Finiteness of the Four Loop N=8 Supergravity,''
  JHEP {\bf 0909} (2009) 116
  [arXiv:0906.3495 [hep-th]].
  %%CITATION = JHEPA,0909,116;%%    




\end{thebibliography}
\end{document}